\documentclass[oldversion]{aa} % for the letters 
\usepackage{graphicx}
\usepackage{txfonts}
\usepackage{natbib}
\bibpunct{(}{)}{;}{a}{}{,}

\pdfoutput=0

\begin{document}
 
\title{The moderate magnetic field on the flare star Proxima Centauri}

\author{A.~Reiners
  \inst{1}\fnmsep\thanks{Emmy Noether Fellow}
 \and
 G.~Basri\inst{2}
}

%\offprints{A. Reiners}

\institute{
  Universit\"at G\"ottingen, Institut f\"ur Astrophysik, Friedrich-Hund-Platz 1, D-37077 G\"ottingen, Germany\\
  \email{Ansgar.Reiners@phys.uni-goettingen.de}
  \and
  Astronomy Department, University of California, Berkeley, CA, 94720, USA\\
  \email{basri@berkeley.edu}
}

\date{Accepted Aug 21, 2008}

% \abstract{}{}{}{}{} 
% 5 {} token are mandatory

\abstract {We report moderate magnetic flux of $450$\,G$\,< Bf <
  750$\,G (3$\sigma$) on the nearby M5.5 flare star Proxima Centauri.
  A high resolution UVES spectrum was used to measure magnetic flux
  from Zeeman broadening in absorption lines of molecular FeH around
  1\,$\mu$m.  The magnetic flux we find is relatively weak compared
  with classical strong flare stars, but so are Proxima’s flaring
  rates and actual emission levels. We compare what is known about the
  rotation rate, Rossby number, and activity levels in this star to
  relations between these quantities that have been recently being
  developed more generally for M dwarfs. We conclude that the magnetic
  flux is higher than the best estimates of the Rossby number from
  period measurements. On the other hand, the activity levels on
  Proxima Centauri are at the high end of what could be expected based
  on the measured field, but not so high as to exceed the natural
  scatter in these relations (other stars lie along this high envelope
  as well).  }


\maketitle
% 
% ________________________________________________________________

\section{Introduction}
\label{sect:Introduction}

Proxima Centauri ($\alpha$ Cen C, GJ~551\,C, in the following
Prox~Cen), our closest stellar neighbor, has been known to be a flare
star for more than 50 years \citep{Thackeray50}. Prox~Cen is of
particular importance for our understanding of very cool stars,
because we can observe it with very high precision. On Prox~Cen,
activity can be detected at very low level, and great effort has been
put into monitoring campaigns to investigate it.  At a spectral type
of M5.5, Prox~Cen is probably completely convective.  It shows
persistent H$\alpha$ emission and flaring, which may lead to the
expectation of rapid rotation ($\ga 3$\,km\,s$^{-1}$). However, the
strength of quiescent activity in Prox~Cen is relatively low compared
with other active mid-M stars \citep{Mathioudakis91, Patten94,
  Hawley96, Bochanski07}, and the flare rate is as well
\citep{Walker81}.

The relation between rotation and magnetic activity is well known in
solar-type stars \citep[e.g.][]{Pizzolato03}. It is observed in very
slowly rotating early-M dwarfs \citep[$v\,\sin{i} \ga
1$\,km\,s$^{-1}$,][]{Reiners07}, too. \citet{Mohanty03} found that the
rotation-acitivity relation is valid in low-mass stars down to
spectral class M9, at least to the extent that rapid rotators show
activity saturation. In solar-type stars, chromospheric and coronal
activity is believed to be due to magnetic fields generated through a
magnetic dynamo, and \citet{RB07} found that magnetic flux scales with
H$\alpha$ activity also in the late-M dwarfs. This all leads to the
expectation that activity on a flare star like Prox~Cen is due to
strong magnetic fields generated through a magnetic dynamo that may be
driven by rotation.

There has been considerable effort to determine the rotation period of
Prox~Cen. \citet{Benedict98} searched for a rotation period in HST FGS
data claiming a rotation period of $P = 83.5$\,d, and \citet{Guinan96}
report a period of 31.5\,d from IUE data. From the intensity of Mg\,II
emission lines in the same IUE data, \citet{Doyle87} predicted a
rotation period around 51\,d according to the rotation-activity
relation. However, \citet{Doyle87} also predicted a rotation period of
27~d for UV~Cet. UV~Cet has a projected rotational velocity of
$v\,\sin{i} = 32.5$\,km\,s$^{-1}$ \citep{Mohanty03}, i.e. a period of
less than about half a day, which means that an estimate for the
rotational period from Mg\,II data is probably not very useful in
mid-M dwarfs. Recently, \citet{Kiraga07} reported a rotation period of
82.5\,d for Prox~Cen, very close to the 83.5\,d period reported by
\citet{Benedict98}. Altogether, there is evidence that $\sim 83$\,d is
the true rotational period of Prox~Cen.

The rotation velocity of Prox~Cen is probably rather low, but still it
shows persistent flaring activity. Although its flare rate is below
those of the very active flare stars like CN~Leo, it is still an
active star and strong magnetic flux might be expected.  In this paper
we present the first direct measurement of the magnetic flux on
Prox~Cen.

\section{Data}

We retrieved our data from the ESO/ST-ECF Science Archive. An
observation of Prox~Cen was taken with UVES at VLT's UT2 on April 18,
2006. The data is described in more detail in \citet{Kurosawa06}; they
were taken with the red arm centered at 860~nm providing wavelength
coverage from 6700--10\,250\,\AA. At a slit length of 1'' the
resolving power is $R \approx 40\,000$. Pipeline-reduced data were
taken from the archive and used for our analysis. Standard reduction
steps involve bias and background subtraction, flat-fielding, optimum
extraction, and order-merging.

The red setting used in this exposure covers the absorption band of
the molecule FeH, which we use for our determination of the magnetic
flux in the following. After an exposure time of 100\,s, a
signal-to-noise ratio of about 60 is achieved in this spectral range.
Unfortunately, H$\alpha$ is not covered by this instrumental setup.

\section{Method}

The method we employ to measure the magnetic flux in Prox~Cen was
introduced by us in \citet{Reiners06} and after that applied to a
sample of M2--M9 dwarfs in \citet{RB07}. Here, we give a brief
overview of the method and refer to these two papers for a more
detailed description.

The absorption band of FeH contains a forest of strong, well isolated
lines of which some are sensitive to the Zeeman effect caused by
magnetic fields, while others are not. In a relatively small spectral
range, we find spectral lines of the same ro-vibrational transition
that react differently to the presence of a magnetic field. The direct
simulation of the Zeeman effect in FeH calculating polarized radiative
transfer is still hampered by the lack of Land\'e factors. Instead, we
choose a more empirical approach: We observed two spectra of early-M
dwarfs for which magnetic field measurements from atomic absorption
lines exist. One of them shows no signs of magnetic fields (and no
activity; GJ~1002), for the other a total magnetic flux of $Bf \sim
3.9$\,kG was measured \citep[Gl~873,][]{JKV00}. The strategy we follow
to measure the magnetic flux in Prox~Cen is the following. We apply an
optical-depth scaling to the two reference spectra so that the
strength of the FeH band matches the strength of FeH absorption in
Prox~Cen. We can now use the magnetically insensitive lines to fix the
rotational velocity, and the magnetically sensitive lines to adjust
for the magnetic flux. This is done by interpolating the two template
spectra (between zero magnetic flux and $Bf =3.9$\,kG) in order to
achieve the best fit to the spectrum of Prox~Cen. Hereby, we assume
that the Zeeman broadening is linear in $Bf$ and that the distribution
of magnetic field strength in Prox~Cen is similar to the distribution
in the magnetic reference star.

\section{Results}

\begin{figure*}
  \resizebox{.5\hsize}{!}{\includegraphics[]{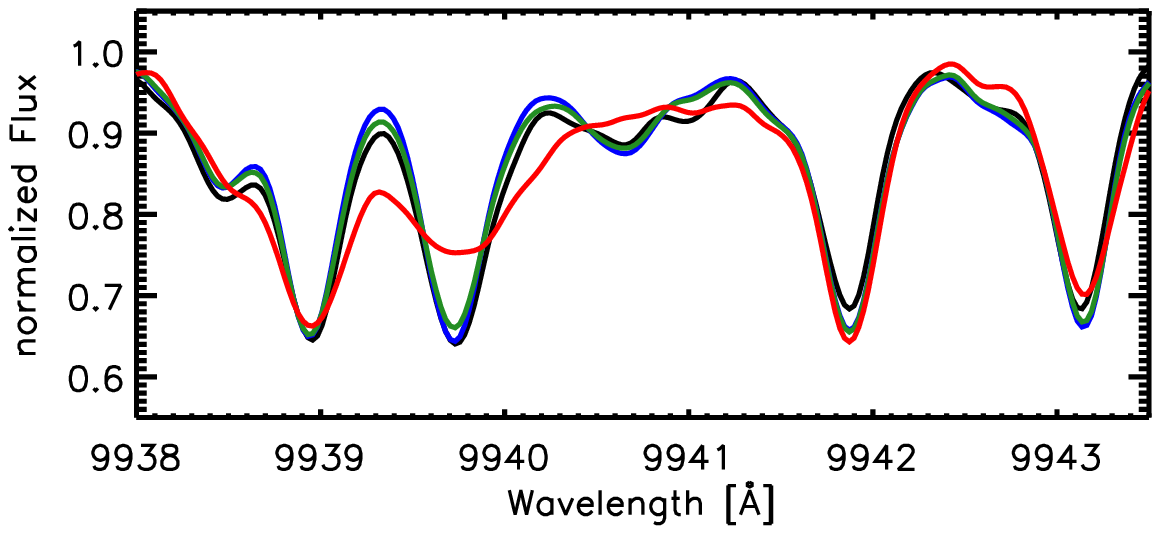}}\\[2mm]
  \resizebox{.9\hsize}{!}{\includegraphics[]{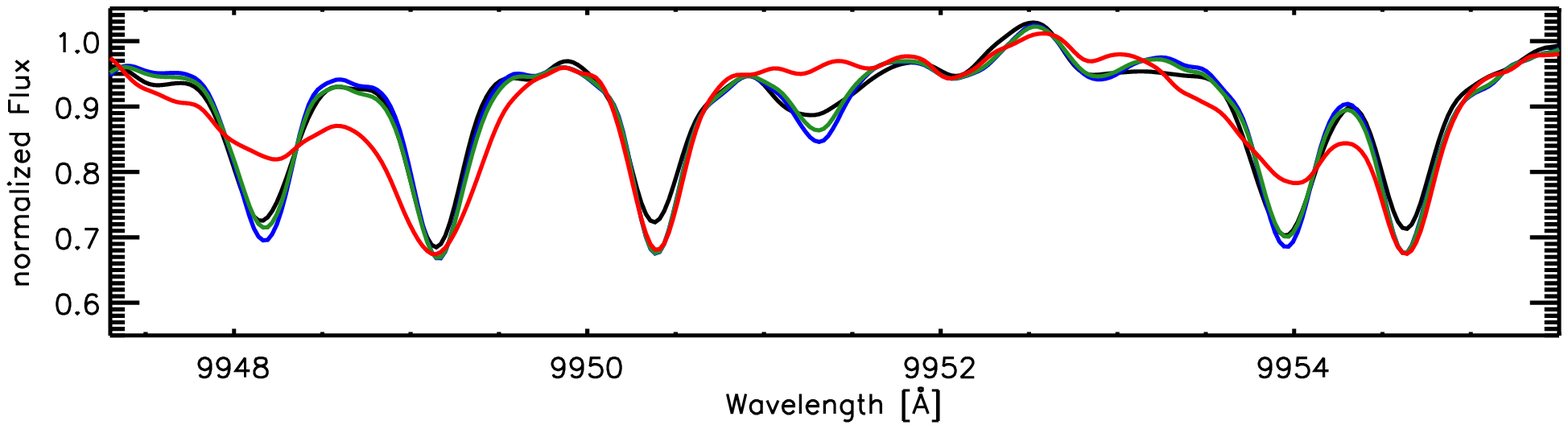}}\\[2mm]
  \resizebox{.9\hsize}{!}{\includegraphics[]{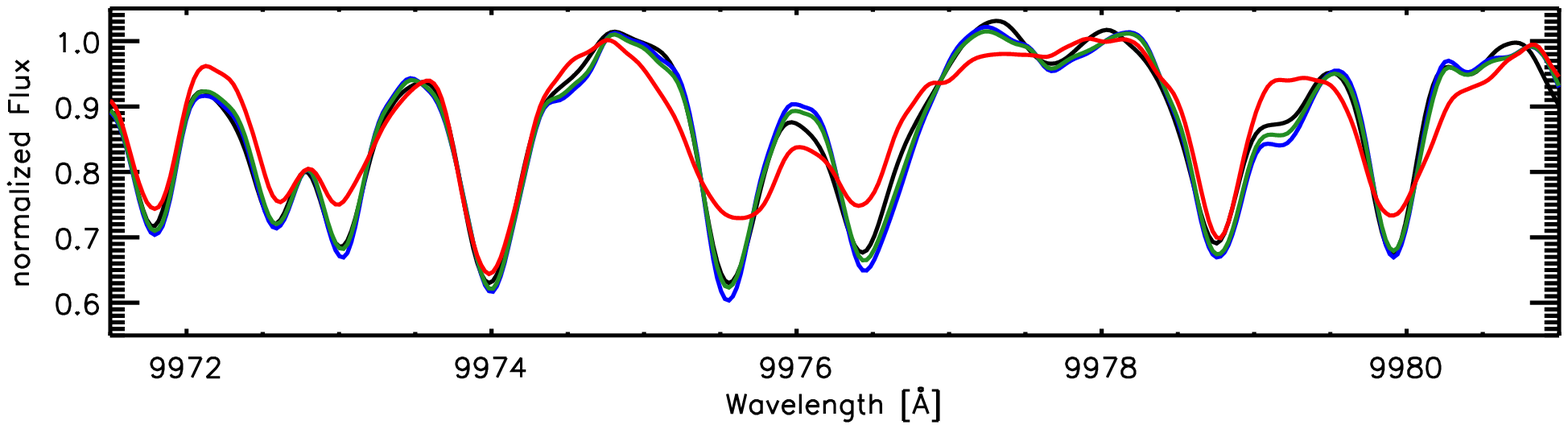}}
  \caption{\label{fig:Fit}Data of Prox~Cen in the wavelength regions
    used for the fit. Black: data of Prox~Cen; red: spectrum of a star
    with strong magnetic flux (3.9\,kG) scaled and artificially
    broadened to match the spectrum of Prox~Cen; blue: as the red line
    but for a star with no magnetic flux; green: best fit linear
    interpolation between the case of very strong and no magnetic flux
    (at a ratio of $\sim$15:85, i.e. $\approx 0.6$\,kG).}
\end{figure*}

Before the magnetic flux can be measured from the interpolation of the
two reference spectra, the latter were adjusted for line depth and
rotational broadening. The strength of the FeH band is about 85\,\% of
the depth seen in GJ~1002 \citep[$a=1.2$, see][]{RB07}, this is
consistent with other old M5.5 dwarfs. The magnetically insensitive
lines in Prox~Cen show no extra broadening with respect to the
template spectra, which implies that the projected rotational velocity
is below the detection limit of $\approx 3$\,km\,s$^{-1}$. This comes
as no surprise, because several long rotational periods were reported
from photometric observations. However, as we explained in
Sect.\,\ref{sect:Introduction}, there is still some uncertainty about
the rotation of Prox~Cen, and to our knowledge no measurement of
$v\,\sin{i}$ was reported for Prox~Cen before.

In Fig.\,\ref{fig:Fit}, we show the spectrum of Prox~Cen (artificially
broadened to match the spectral resolution of the template spectra)
together with the reference spectra of GJ~1002 and Gl~873 scaled in
optical depth to match the line strengths of Prox~Cen. We searched for
the best fit of the interpolation between the two reference spectra
and the spectrum of Prox~Cen calculating the summed quadratic
difference $\chi^2$. The fit was calculated in three wavelength
ranges, 9938.0--9943.5\,\AA, 9947.3--9955.5\,\AA, and
9971.5--9981.0\,\AA; 901 pixels are used altogether. All three regions
are shown in Fig.\,\ref{fig:Fit}. The best fit is achieved for a ratio
between our spectra with strong magnetic flux to zero flux of 15:85,
i.e. $Bf = 0.15 \times 3.9$\,kG~$\approx 600$\,G. With a SNR of 60, we achieve
a reduced chi-square of $\chi^2_{\nu} = 0.86$ indicating a somewhat
underestimated SNR ($\chi^2_{\rm min} = 775$).

\begin{figure}
  \center
  \resizebox{.9\hsize}{!}{\includegraphics[]{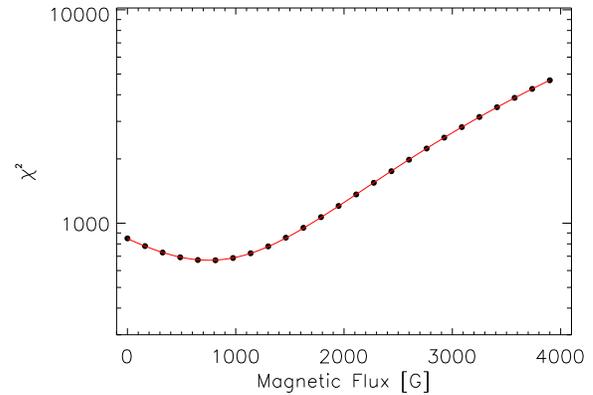}}
  \caption{Chisquare as a function of magnetic flux (see
    text).\label{fig:Chisquare}}
\end{figure}

\begin{figure}
  \center
  \resizebox{.9\hsize}{!}{\includegraphics[]{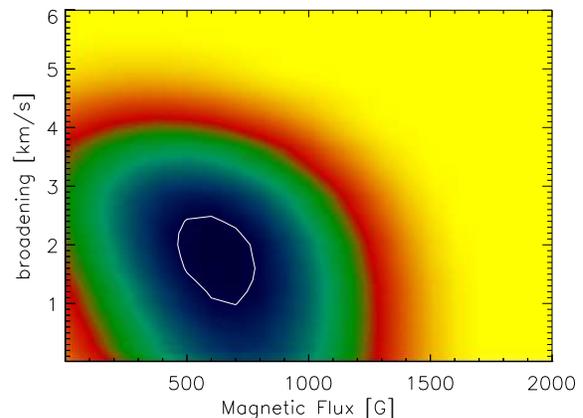}}
  \caption{Chisquare landscape as a function of $Bf$ and $v\sin{i}$.
    The white line marks the region where $\chi^2 < \chi^2_{\rm min} +
    9$. \label{fig:landscape}}
\end{figure}

The fit quality in terms of $\chi^2$ is shown as a function of
magnetic flux $Bf$ in Fig.\,\ref{fig:Chisquare}. We compute a formal
3$\sigma$ uncertainty in $Bf$ from searching the range of $Bf$ for
which $\chi^2 < \chi^2_{\rm min} + 9$ while varying all other
parameters (including rotational broadening). The formal result of
this exercise is:
\begin{equation}
  Bf = 600~\mathrm{G} \pm 150~\mathrm{G~(3\sigma)}
\end{equation}

There is some degeneracy between the derived magnetic flux and the
rotational broadening $v\,\sin{i}$. In principle, we could use the
magnetically insensitive lines to determine the broadening that
applies to all spectral lines, i.e., residual instrumental broadening
or $v\,\sin{i}$. With this information, we could then determine
magnetic broadening using the magnetically sensitive lines. This
approach has to caveats: I) Lacking molecular Land\'e factors, we
cannot entirely trust the assumption that magnetically ``insensitive''
lines show absolutely no effect to magnetic fields -- small broadening
due to magnetic fields may occur even in those lines; II) Using
subsets of the spectrum would lead to different estimators of fit
quality that are not necessarily comparable. This means that the
parameters providing high fit quality in the magnetically insensitive
lines may not lead to a good fit in the magnetically sensitive lines
if only the magnetic flux is changed.  This can only be avoided if all
lines are used simultaneously. Thus, we prefer to always compute our
$\chi^2$ values from the full spectral range.

In order to investigate the crosstalk between $Bf$ and $v\sin{i}$ (or
any instrumental broadening), we provide the landscape of $\chi^2$ as
a function of overall broadening (e.g., $v\,\sin{i}$) and $Bf$ in
Fig.\,\ref{fig:landscape} This landscape shows the clear minimum in
$\chi^2$ around $Bf \approx 600$\,G. The minimum is also somewhat
dependent on the choice of $v\,\sin{i}$.  Interestingly, very slow
values of ``$v\,\sin{i}$'' are excluded. As mentioned above, the
detection limit of overall line broadening at the spectral resolution
of our data is on the order of 3\,km\,s$^{-1}$, i.e.  anything below
3\,km\,s$^{-1}$ is not necessarily due to rotation, but is within the
uncertainties from instrumental broadening. In other words, the fact
that we require a broadening on the order of 2\,km\,s$^{-1}$ to fit
the data does not imply that Prox~Cen is rotating at this level,
because at the given resolution a formal result of $v\,\sin{i} =
2$\,km\,s$^{-1}$ is fully consistent with zero rotation. On the other
hand, the detection of magnetic flux Bf is from a differential
analysis comparing lines of different magnetic sensitivity. Hence no
absolute lower limit of detectability (other than the calculated
uncertainty that also depends on spectral resolution) applies.

\section{Discussion}

\subsection{Activity and Magnetic Flux}

Active M dwarfs typically exhibit magnetic flux of a few kG
\citep{RB07}. However, the magnetic flux on Prox~Cen is substantially
lower than 1\,kG although it clearly is an ``active'' star.  Does
Prox~Cen have particularly weak magnetism given its level of magnetic
activity?

The flare rate of Prox~Cen is comparably low; \citet{Walker81} report
35 Flares during 25.25\,h, which means it is less actively flaring
than UV~Cet, CN~Leo, or EQ~Peg. In particular, the flare rate of $\sim
1.4$\,h$^{-1}$ found by \citet{Walker81} is only about 60\,\% the
flare rate of CN~Leo found by \citet{Kunkel73} at comparable U-band
sensitivity. This indicates that Prox~Cen may be a flare star with
relatively little activity.

Several measurements of H$\alpha$ emission on Prox~Cen are reported in
the literature. \citet{Mathioudakis91} measured an H$\alpha$
equivalent width of 1.67\,\AA. \citet{Patten94} report a minimum of
3\,\AA, and 27\,\AA\ during flare state, and \citet{Hawley96} found an
H$\alpha$ emission line of 3.01\,\AA\ equivalent width. The mean
H$\alpha$ equivalent width in active M5 dwarfs is $5.9 \pm 1.1$\,\AA\,
\citep[][although one has to keep in mind that this is from data of
lower resolution]{Bochanski07}. The normalized H$\alpha$ luminosities
corresponding to H$\alpha$ equivalent widths of 1.67\,\AA\ and
3.0\,\AA\ are $\log{L_{{\rm H}\alpha}/L_{\rm bol}} = -4.2$ and $-4.0$,
respectively \citep[see][]{RB08a}.\footnote{\citet{Hawley96} calculate
  a value of $\log{L_{{\rm H}\alpha}/L_{\rm bol}} = -4.2$ for an
  equivalent width of 3.0\,\AA, i.e., 0.2 dex lower than our
  calibration.}  According to \citet{Mohanty03}, $\log{L_{{\rm
      H}\alpha}/L_{\rm bol}} \approx -4.2$ is the level where activity
saturates in stars of spectral type M5.5--M8.5, i.e., activity in
Prox~Cen is barely in the regime of saturation. In fact, the lowest
value of H$\alpha$ emission (1.67\,\AA) is comparable to H$\alpha$
measured in GJ\,1286 (1.50\,\AA) by \citet{RB07}.  This M5.5 star
exhibits magnetic flux of $0.4 \pm 0.2$\,kG (1$\sigma$), which is very
similar to Prox~Cen (and GJ~1286 is a slow rotator).

Another proxy of magnetic activity are X-rays. X-ray emission of
Prox~Cen was measured with ROSAT and XMM: The X-ray flux measured with
ROSAT is $\log{L_{\rm X}} = 27.2$\,erg\,s$^{-1}$ \citep{Hunsch99}.  At
$M_{\rm bol} = 11.98$ \citep{Hawley96} this means that the normalized
X-ray luminosity is $\log{L_{\rm X}/L_{\rm bol}} = -3.5$.  Newer
observations from XMM are consistent with this result
\citep[see][]{NEXXUS}. \citet{Kiraga07} report a normalized X-ray
luminosity of $\log{L_{\rm X}/L_{\rm bol}} = -3.8$. This value
probably comes from the two longest ROSAT PSPC observations only, for
which $\log{L_{\rm X}} \approx 26.9$ is provided by \citet{NEXXUS}
(while $\log{L_{\rm X}} \approx 27.4$ is given for the two XMM
exposures).

For a mid-M dwarf, a normalized X-ray luminosity between $\log{L_{\rm
    X}/L_{\rm bol}} = -3.5$ and $-3.8$ is consistent with a normalized
H$\alpha$ luminosity between $-3.7$ and $-4.2$ \citep[although the
scatter is fairly large, see][]{RB07}. This is consistent with
H$\alpha$ measurements (see above).  Both, X-ray- and H$\alpha$
emission put Prox~Cen on the low end of active M dwarfs.

\citet{RB07} investigated the relation between magnetic flux and
H$\alpha$ emission among M dwarfs. From their sample, a star of
spectral class M5.5 with a normalized H$\alpha$ luminosity of
$\log{L_{{\rm H}\alpha}/L_{\rm bol}} \approx -4$ can be expected to
have a magnetic flux level on the order of $Bf \approx 1$\,kG.  For
$Bf \approx 600$\,G, i.e. the magnetic flux found on Prox~Cen,
H$\alpha$ emission on the order of $\log{L_{{\rm H}\alpha}/L_{\rm
    bol}} = -4.2$ can be expected. Thus, the magnetic flux estimated
from the lowest X-ray and H$\alpha$ detections is consistent with the
magnetic flux measured here, while the emission seen in all X-ray and
H$\alpha$ data is scattering around a somewhat higher level. This
leads to the conclusion that Prox~Cen has somewhat higher activity
than would be predicted from its magnetic field, but not an
unreasonably higher amount, given that the scatter in all these
relations is substantial.

\subsection{Rotation and Rossby number}

In M5.5--M8.5 stars, the rotation-activity relation saturates at a
surface velocity of about 10\,km\,s$^{-1}$ \citep{Mohanty03}.
However, there are also slowly rotating mid-M stars that exhibit
strong magnetic activity. While rapid rotators always seem to be
active, the opposite is not true. Prox~Cen shows no detectable Doppler
broadening due to rotation, its projected rotation velocity is smaller
than $v\,sin{i} \approx 3$\,km\,s$^{-1}$.  \citet{RB07} found that all
M dwarfs with magnetic flux $Bf < 1$\,kG are slow rotators
($v\,\sin{i} < 3$\,km\,s$^{-1}$). The low magnetic flux found on
Prox~Cen is consistent with slow rotation.

At a radius of $R=0.145$\,R$_{\sun}$ \citep{Segransan03}, the
spectroscopic limit on the rotation velocity translates into a
projected rotation period of $P/\sin{i} > 1$\,d.  There is growing
evidence that the rotation period of Prox~Cen is on the order of 80\,d
\citep{Benedict98, Kiraga07}, which is consistent with the lack of
Doppler broadening (although it is almost two orders of magnitude
below the detection limit). That long a rotation period would imply an
extremely low surface rotation velocity of $v = 90$\,m\,s$^{-1}$.

The turnover time of Prox~Cen is in the range 70--100\,d
\citep{Gilliland86}. If the rotation period of Prox~Cen is on the
order of 80\,d, its Rossby number is on the order of $Ro = 1$ (or
$\log{Ro} = 0$). This high a Rossby number would put Prox~Cen among
the stars occupying the very beginning of the linear part of the
rotation-activity relation, i.e. among the stars with the lowest
measurable activity. From the relation between Rossby number and
magnetic flux \citep{Saar01, RB08b} one would expect magnetic flux on
the order of 100\,G for such a small Rossby number. Normalized X-ray
activity would be on the order of $\log{L_{\rm X}/L_{\rm bol}} = -5.0$
\citep{Pizzolato03, Kiraga07}, and H$\alpha$ activity also would be at
about $\log{L_{{\rm H}\alpha}/L_{\rm bol}} \approx -5$ implying that
H$\alpha$ would probably not be detectable as an emission line. The
conclusion from the various rotation-activity relations is that from
H$\alpha$ emission, X-ray emission, and from magnetic flux, one would
estimate a Rossby number on the order of $\log{Ro} \approx -0.6$.
Together with a convective overturn time of $\tau_{\rm conv} =
70..100$\,d, this leads to a rotation period estimate between 17\,d
and 25\,d rather than 80\,d. To reach $\log{Ro} \approx -0.6$ at a
rotation period of 80\,d, the turnover time needs to be on the order
of 300\,d rather than 100\,d.

We conclude that Prox~Cen is a slow rotator with a rotation period
longer than about 15\,days, but its activity and magnetic flux are
indicative of a rotation period substantially shorter than 80\,d. It
is less active than more rapid rotators like CN~Leo or UV~Cet. Its
level of chromospheric and coronal activity is relatively high for its
moderate magnetic flux or Rossby number, but given the substantial
scatter in the relations between these quantities, it merely lies with
other stars on the high activity side of the linear relation between
rotation, activity, and magnetic flux in M dwarfs.

\begin{acknowledgements}
  Based on observations made with the European Southern Observatory
  telescopes obtained from the ESO/ST-ECF Science Archive Facility.
  A.R.  acknowledges research funding from the DFG as an Emmy Noether
  fellow under RE 1664/4-1. G.B. acknowledges support from the NSF
  through grant AST-0606748.
\end{acknowledgements}

\end{document}